\documentclass[twocolumn,pr,showpacs,aps,superscriptaddress]{revtex4}
\usepackage{multirow}
\usepackage{graphicx} 
\usepackage{color}

\begin{document}
\bibliographystyle{apsrev}
\title{Origin of the Fermi arcs in cuprates: a dual role of \\ quasiparticle and pair excitations}

\author{William Sacks}

\author{Alain Mauger}

\affiliation{Institut de Min\'{e}ralogie, de Physique des Mat\'{e}riaux et
de Cosmochimie, CNRS, UMR 7590}

\author{Yves Noat}

\affiliation{Institut des Nanosciences de Paris, CNRS, UMR 7588 \\
Sorbonne Universit\'{e}, Facult\'{e} des Sciences et Ing\'{e}nierie, 4 place
Jussieu, 75005 Paris, France}

\date{\today}



\begin{abstract}
ARPES mesurements in cuprates have given key information on the
temperature and angle dependence of the gap ($d$-wave order
parameter, Fermi arcs and pseudogap). We show that these features
can be understood in terms of a Bose condensation of interacting
{\it pairons} (preformed hole pairs which form in their local
antiferromagnetic environment). Starting from the basic properties
of the pairon wavefunction, we derive the corresponding $k$-space
spectral function. The latter explains the variation of the ARPES
spectra as a function of temperature and angle up to $T^*$, the
onset temperature of pairon formation. While Bose excitations
dominate at the antinode, the fermion excitations dominate around
the nodal direction, giving rise to the Fermi arcs at finite
temperature. This dual role is the key feature distinguishing
cuprate from conventional superconductivity.
\end{abstract}

\maketitle

{\it Introduction}.

\vskip 2mm

More than thirty years after their discovery by G. Bednorz and A.
M\"{u}ller \cite{ZPhys_Bednorz1986}, cuprates still keep their secret.
The parent compounds are two-dimensional insulating antiferromagnets
and superconductivity is induced by doping the CuO planes. The
origin of the superconducting $T_c$ dome is still unknown and the
Bardeen-Cooper-Schrieffer theory (BCS) \cite{PR_BCS1957} fails to
describe the essential properties of cuprates :\,the pairing and
condensation mechanisms.

In addition to tunneling spectroscopy \cite{Revmod_Fisher2007},
angle resolved photoemission spectroscopy (ARPES) has revealed
ground-breaking information on the superconducting condensate, in
particular the angular dependence, which otherwise must be inferred
(see for instance
\cite{Revmod_Damascelli2003,NatPhys_Hashimoto2014,LTPhys_Kordyuk2015}
and refs. therein). There are three major empirical findings\,: the
$d$-wave nature of the condensate, a partial Fermi-surface at finite
temperature (Fermi arcs) \cite{Nat_Norman1998} and a spectral gap
which persists at $T_c$ in the antinodal direction up to a higher
temperature $T^*$. The latter {\it pseudogap}
\cite{Rep_ProgPhys_Timusk1999}, first found by NMR spectroscopy
\cite{PRL_Alloul1989,PRL_Warren1989}, was then observed in the
specific heat of YBa$_2$Cu$_3$0$_{6+x}$ \cite{PhysicaC_Loram1994}
and subsequently by ARPES \cite{Sci_Loeser_1996,Nat_Ding1996} and
tunneling \cite{PRL_renner1998_T}. The relation between the
pseudogap state and non-BCS superconductivity is still strongly
debated.

In earlier works \cite{Nat_Norman1998}, it was suggested that the
Fermi arcs seen above the critical temperature collapse to virtually
a single point in $k$-space at the nodes at $T_c$, compatible with a
$d$-wave symmetry of the order parameter. However, more recent data
with better resolution
\cite{NatPhys_Hashimoto2014,PRB_Ideda2012,PhilMag_Kaminski2015} show
that a finite-sized Fermi arc still exists at $T_c$ around each
node. Its origin is undecided and it is unclear whether these Fermi
arcs are tied to $T_c$ or rather to $T^*$. Recent effort has focused
on the gap function near the node (nodal gap), often proposed to be
the superconducting order parameter
\cite{PRB_Ideda2012,ncomm_Anzai}. In this context, the antinodal
pseudogap is attributed to some extraneous competing order, such as
a spin or charge density wave \cite{Sci_Tanaka2006,PNAS_Vishik2012}.
On the other hand, based on ARPES measurements in the antinodal
direction, A. Kanigel et al. \cite{PRL_kanigel} and  M. Shi et al.
\cite{EPL_shi} stressed the preformed-pair interpretation.

In this article, we answer these questions in the framework of the
condensation of preformed {\it pairons} \cite{EPL_Sacks2017}. To
proceed, we calculate the spectral function for cuprates and
directly compare the computed energy distribution curves (EDCs) to
the ARPES measurements as a function of temperature and angle at the
Fermi surface. We show that the condensation of pairons fully
describes the continuous evolution of the ARPES spectra with
temperature and angle, from the antinode to the node. Remarkably,
the boson condensation is revealed in the antinodal (AN) direction
where Bose-Einstein statistics dominate, whereas the fermion
excitations dominate near the nodal (N) direction. At $T_c$ and
above, the pseudogap is revealed by the incoherent pair excitations
at the antinode, which coexist with the Fermi arcs around the nodes
-- a direct consequence of the composite character of pairons.


\vskip 2mm {\it From pairons to Cooper pairs}. \vskip 2mm

In the framework of the $t-J$ Hamiltonian, it has been shown that
two holes in an antiferromagnetic system form a bound state provided
that the ratio $J/t$ is sufficiently large
\cite{PRB_Poilblanc1994,RevModPhys_Dagotto1994}. In a recent work
\cite{EPL_Sacks2017}, we have extended this mechanism to a more
realistic system with a large number of holes. In this scenario,
pairs of holes are trapped in their antiferromagnetic environment,
on the scale of the antiferromagnetic correlation length $\xi_{AF}$,
forming {\it pairons}. This idea is strongly supported by the
experimental finding that $\xi_{AF}\sim 1/\sqrt{p}$
\cite{PRL_Ando2001}, where $p$ is the number of holes per copper
atom. Thus $\xi_{AF}$ varies as the distance between holes (or hole
pairs), providing an immediate explanation for the linear variation
of the antinodal gap, $\Delta_p$, with hole doping $p$
\cite{EPL_Sacks2017}.


We start with a boson Hamiltonian corresponding to a gas of
non-interacting pairons which, in absence of mutual interactions,
describes the incoherent pseudogap state\,:
\begin{equation}
H_B=\sum_{i}\varepsilon_i\,b_i^\dag b_i \ ,
\label{Hbosons}
\end{equation}
where the operator $b_i^\dag$ creates a given boson of energy
$\varepsilon_i$.

\begin{figure}
\centering \vbox to 3.5 cm{
\includegraphics[width=7 cm]{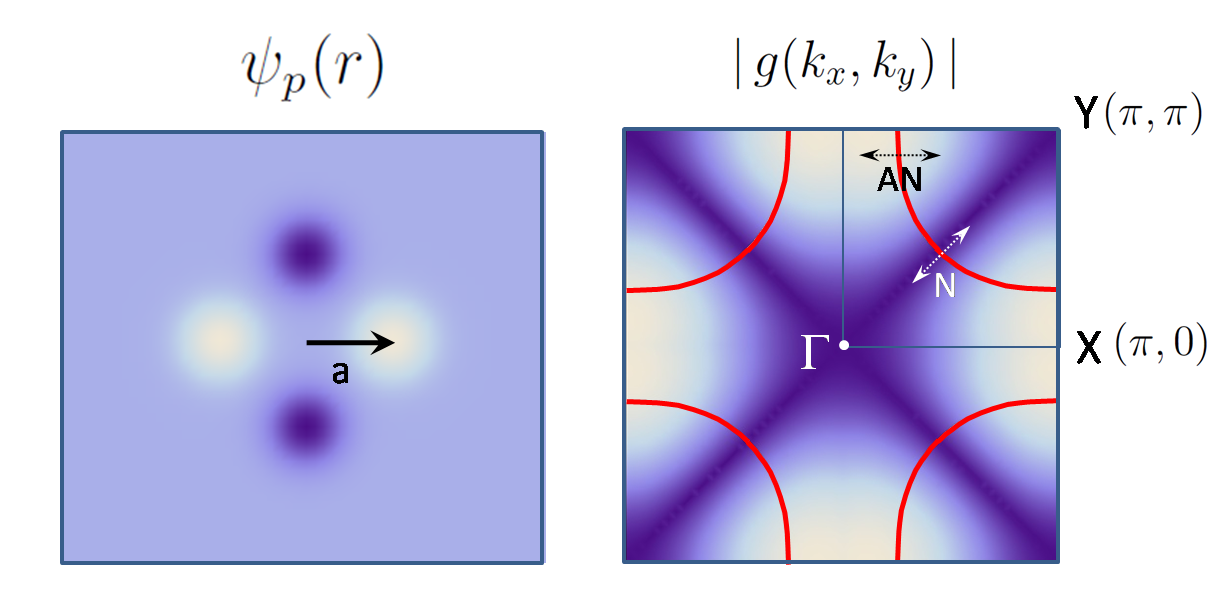}}
\caption{(color online) Pairon wavefunction (left panel) in the
center of mass ($\vec R = 0$) and its Fourier transform (right
panel). The Fermi surface, indicated by the red curve in the right
panel, is calculated based on the accurate band structure proposed
by Markiewicz et al. \cite{PRB_Markiewicz2005}. In the right panel
the antinodal direction (AN or $\theta=0$) and the nodal direction
(N or $\theta = \pi/4$) are indicated.}\label{Wave_fct}
\end{figure}

In the two-particle center of mass the pairon wavefunction
$\psi_i(\vec{r})$, (where $\vec{r}=\vec{r_1}-\vec{r_2}$ is the
distance between the two holes) is determined by the AF environment,
which imposes its symmetry. As a result, $\psi_i(\vec{r})$ has to
vanish along the lattice diagonal, which corresponds to the nodal
direction in $k$-space (see Fig.\ref{Wave_fct}). We thus take the
wavefunction to be the generic form\,:
\begin{eqnarray}\label{mfeq5}
 \psi(\vec{r})& = & \frac{1}{\sqrt{4}}\left[\varphi(\vec{r}-a\,\hat{x})+\varphi(\vec{r}+a\,\hat{x})\right. \\
&-&
\left.\varphi(\vec{r}-a\,\hat{y})-\varphi(\vec{r}+a\,\hat{y})\right]
\ , \nonumber
\end{eqnarray}
where $a$ is the lattice parameter, and
$\varphi(\vec{r})={e^{-\frac{{r}^2}{2b^2}}}/{\sqrt{2\pi b^2}}$. The
parameter $b$ fixes the spatial extension of the pairon wavefunction $\psi_i$ (see Fig.\ref{Wave_fct}).

The pairon wavefunction $\psi_i$ and associated
operator $b_i$ can equivalently be described by a superposition of delocalized
Cooper pairs,
\begin{equation}
b_i=\sum_{\vec{k}}\,g_{\vec{k}}^i\,b_{\vec{k}}^i \ ,
\end{equation}
where the operator
$b_{\vec{k}}^{i\dag}=c_{\vec{k}}^{i\dag}c_{-{\vec{k}}}^{i\dag}$
creates a Cooper pair
$\left|\vec{k}\uparrow,-\vec{k}\downarrow\right\rangle$. In this
formulation, just as in the original Cooper-pair problem
\cite{PR_BCS1957}, the ground state of the system is constructed
from pairs in the zero-momentum state. The weight $g_{\vec{k}}$
appearing in the sum is given by the Fourier transform of the
wavefunction $g^i(k_x,k_y)=\int
e^{i\vec{k}.\vec{r}}\psi_i(\vec{r})d^2\vec{r}$.

Taking the quantum average\,: $b_i^+b_i \approx \left\langle
b_i^+\right\rangle b_i+b_i^+\left\langle b_i \right\rangle$, and
using the standard BCS expression $\left\langle
b_i\right\rangle=\frac{\Delta_k^i}{2E_k^i}$ we obtain the mean-field
Hamiltonian\,:
\begin{equation}
H_{MF}=\sum_{{\vec{k}},i} \epsilon_{\vec{k}}
c_{\vec{k}}^{i+}c_{\vec{k}}^{i}+\sum_{i,{\vec{k}}}\Delta_{\vec{k}}^i
b_{\vec{k}}^{i+}+\sum_{i,{\vec{k}}}\Delta_{\vec{k}}^{i*}
b_{\vec{k}}^i \ ,
\label{H_CPG}
\end{equation}
where the first term is the kinetic-energy and the second is the
pairing term. In the latter, the binding energy $\Delta_k^i$ is
determined by the self-consistent equation
\begin{equation}
\Delta_{\vec{k}}^i=\varepsilon_i\ \sum_{{\vec{k}}'}\
\,g_{\vec{k}}^i\,g_{{\vec{k}}'}^i\
\left(\frac{\Delta_{{\vec{k}}'}^i}{2E_{{\vec{k}}'}^i}\right)\ .
\label{Delta_ki}
\end{equation}
In the continuum limit the sum is replaced by an integral in the
standard fashion. The latter expression bears a strong analogy with
the BCS gap equation, however both the integration limits and the
integrand involving $g_{\vec{k}}$ differ quantitatively.

Dropping the $i$ index, one can show that $g_{\vec{k}}$ takes the
form\,:
\begin{equation}
g_{\vec{k}} \propto e^{-k^2 b^2/2} \left[\cos(k_x a) - \cos(k_y
a)\right]\ ,
\end{equation}
which reveals both the extent of the $k$-states involved, and
$g_{k_F}\propto \cos(2\theta)$ imposes the $d$-wave dependence of
the gap parameter. Note that equations (\ref{H_CPG}) and
(\ref{Delta_ki}) imply the existence of quasiparticles of the
form\,: $$E_{\vec{k}}^i = \sqrt{\epsilon_{\vec{k}}^2 +
{\Delta_{\vec{k}}^i}^2}\ ,$$ associated with degenerate pairons of
binding energy $\varepsilon_i$.

\begin{figure}[t]
\centering \vbox to 9 cm{
\includegraphics[width=7.6 cm]{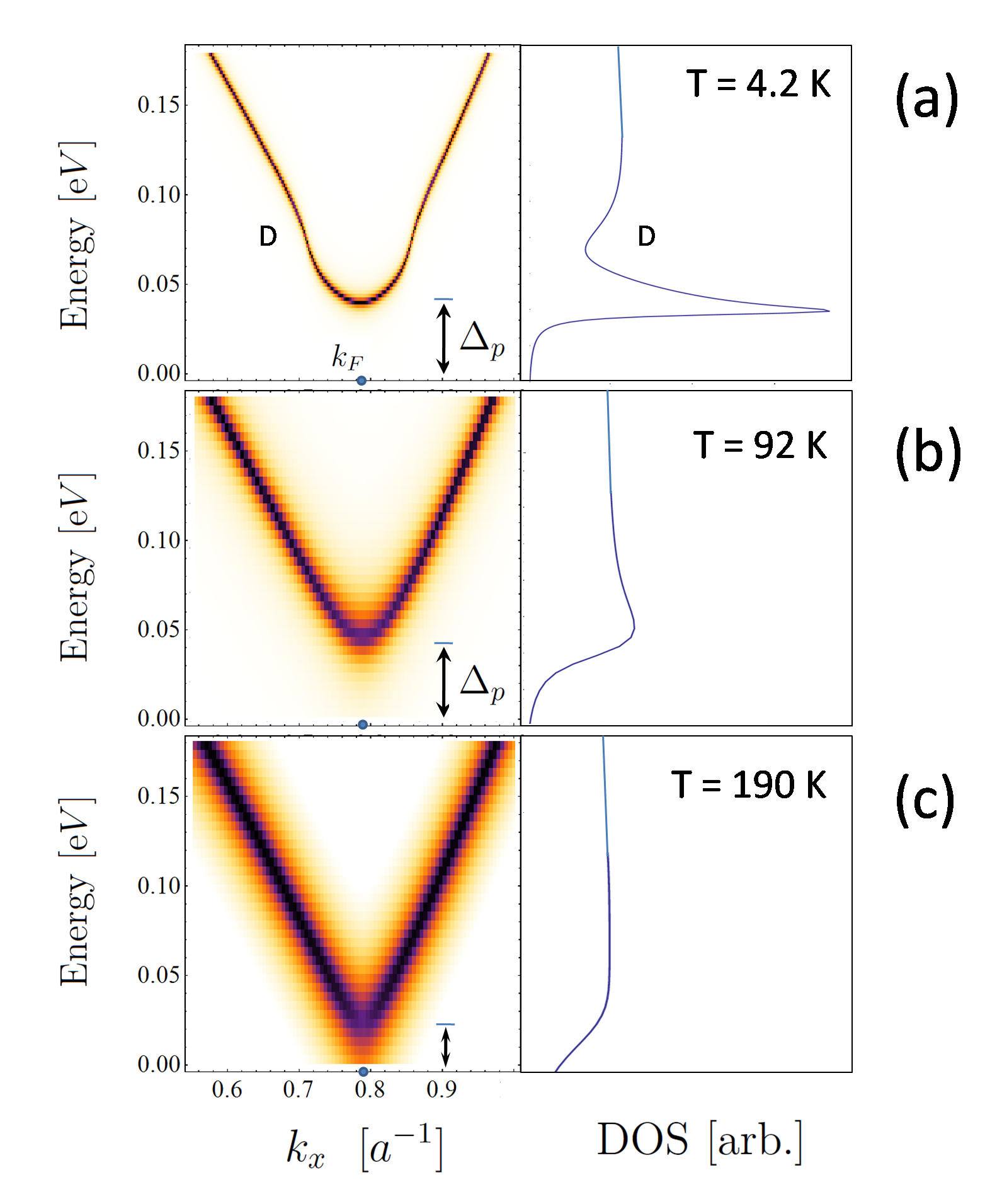}}
\caption{(color online) Spectral function $A(\vec{k},E)$ and
associated DOS along the antinodal direction at low temperature
(panel (a), $T=$4.2K), at the critical temperature (panel (b),
$T=$92K) and close to $T^*$ (panel (c), $T=$190K), with the
broadening parameter $\Gamma$ in Eq.\ref{Spec_fct} given by
$\Gamma=k_BT$. Note that we plot here the excited-state spectral
function where the coherence factors are absent (see
text).}\label{Fig_dos}
\end{figure}

Let us emphasize that the $i$-index is formally equivalent to a band
index in the context of multi-band superconductivity
\cite{PRL_Suhl1959}. However in our case, the $i$-index represents
pairons of differing binding energies and the Hamiltonian
(\ref{H_CPG}) represents a non-superconducting state of independent
Cooper pairs. In order to establish a macroscopic coherent state, a
coupling between pairs of different energies is
necessary\cite{SciTech_Sacks2015, SSC_Sacks2017,JPCM_Sacks2017}.
Without the interaction term between pairons, no long range order is
possible. Still, a gap in the density of states is present at the
Fermi level without the characteristic peaks indicating phase
coherence, see Fig.\,\ref{Fig_dos}(b). Thus the Hamiltonian
(\ref{H_CPG}) provides a description of the main features of the
{\it pseudogap} state.

Our numerical study of the self-consistent equation (\ref{Delta_ki})
shows that, for values of $\varepsilon_i$ in the relevant range for
cuprates ($\varepsilon_i\sim$\,80-210\,meV), $\Delta_{\vec{k}}^i$
can be well approximated by
\begin{equation}
\Delta_{\vec{k}}^i=c\,\varepsilon_i\,g_{\vec{k}}\ ,
\end{equation}
where $c$ is a constant. Taking the parameter $b = a/4$ in the
calculation, we find $c \simeq 0.25$ and an energy gap $\sim
40\,$meV at optimal doping. The satisfying conclusion is that the
Cooper-pair binding energy $\Delta_{\vec{k}}^i$ is proportional to
the pairon energy $\varepsilon_i$, and thus both concepts are
formally equivalent.

In the absence of pairon-pairon interactions, the system is in an
incoherent state, the `Cooper glass' \cite{SciTech_Sacks2015,
SSC_Sacks2017,JPCM_Sacks2017}. The pairon energies are distributed
with a pair density of states $P_0(\Delta_i)$, characterized by the
mean value $\Delta_0$ and dispersion $\sigma_0$. A convenient form
is the Lorentzian\,:
\begin{equation}
P_0(\Delta_i) \propto \frac{\sigma_0^2}{(\Delta_i-\Delta_0)^2 +
\sigma_0^2} \ .\label{P0}
\end{equation}
As will be revealed in the spectral function, Eqs.(\ref{H_CPG} -
\ref{P0}) describe the non-superconducting incoherent state giving
rise to the pseudogap phenomena at the critical temperature and
above.

\vskip 2 mm

{\it Pairon condensation}

\vskip 2 mm

As a result of the interaction between pairons, described by the
additional coupling term
$$H_{int}=\sum_{i\neq j}V_{ij}\,b_i^\dag\,b_j \ ,$$
the system condenses in a homogeneous ground state where pairons lie
in the same quantum state. All Cooper pairs are then characterized
by a unique binding energy $\Delta^i = \Delta_p$. In the mean field
approximation, $\Delta_p$ is given by a self-consistent equation,
which in the antinodal direction has the
form\,\cite{SciTech_Sacks2015}:
\begin{equation}
\Delta_p=\Delta_0-2\beta P_0(\Delta_p) \ . \label{Gap_eq}
\end{equation}
It includes a pair-field term proportional to the average
interaction energy $\beta$. In previous work we showed that this
interaction follows the critical dome and has the value $\beta \sim
2 \,k_B\,T_c$. Moreover, the pairon condensate model matches the
phase diagram for a wide range of doping in terms of a single energy
scale, $J$, the exchange energy \cite{EPL_Sacks2017}.

Using the gap equation (\ref{Gap_eq}), and the energy bands
$\epsilon_k$ of cuprates from Markiewicz et
al.\,\cite{PRB_Markiewicz2005}, the spectral function of
quasiparticles is obtained for any wavevector $\vec{k}$. Assuming
nearly optimal doping ($\Delta_p(0) = 40\,$meV, $T_c = 92\,$K) the
spectral intensity in the AN direction crossing $k_F$ is shown in
Fig.\,\ref{Fig_dos} (left panel) at three different temperatures.
The $k$-sum of these spectra gives directly the associated
quasiparticle density of states (DOS) relevant to tunneling (right
panel).

A well-defined gap is visible at low temperature with a strong
change of slope above the gap energy $\Delta_p$ giving rise to the
characteristic dip in the DOS (right panel a), which has been widely
observed by tunneling spectroscopy in cuprates (see
\cite{Revmod_Fisher2007} and refs. therein). At the critical
temperature (right panel b), a pseudogap persists in the DOS which
finally disappears at a much higher temperature $T^*$ (right panel
c). In the remainder of this work we focus on the EDCs with the wave
vector on the Fermi surface to compare with the experiments.

\begin{figure}
\includegraphics[width=7 cm]{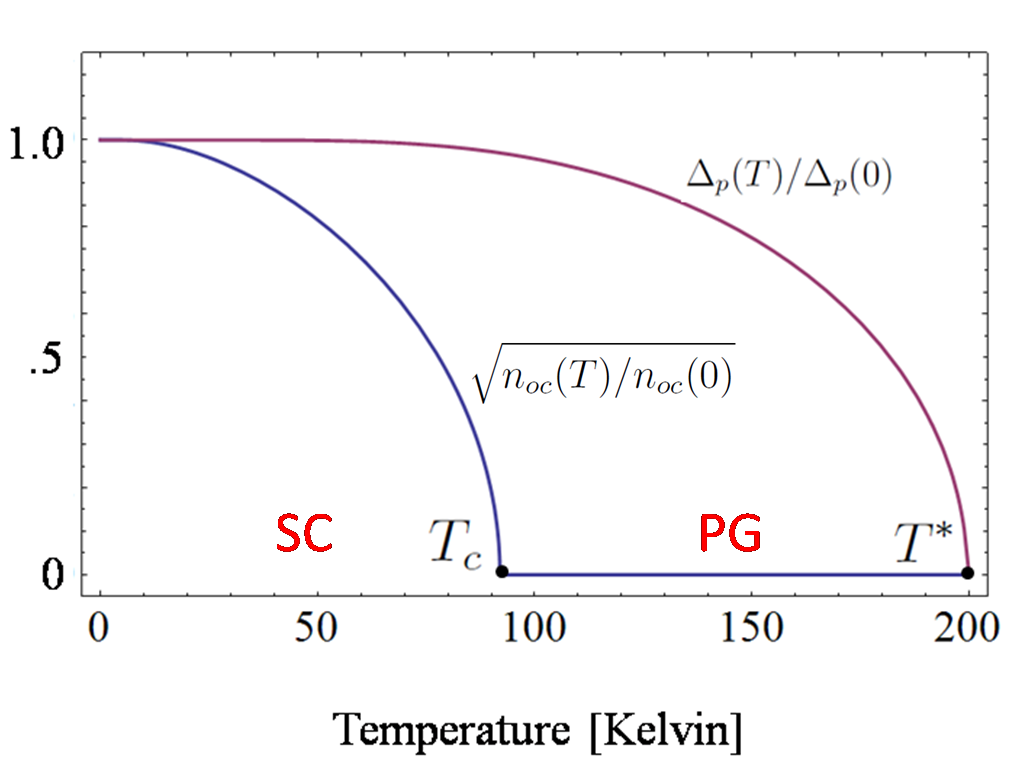}
\caption{(Color online) Temperature phase diagram in the pairon
model for optimal doping\,: critical curves corresponding to
$\sqrt{n_{oc}(T)}$ (lower curve) and the antinodal gap $\Delta_p(T)$
(upper curve). As indicated, the condensate density vanishes at
$T_c$, while pairs exist up to $T^*$.} \label{Fig_noc}
\end{figure}

\vskip 2mm {\it Pair densities}. \vskip 2mm

\begin{figure*}
\includegraphics[scale=.4]{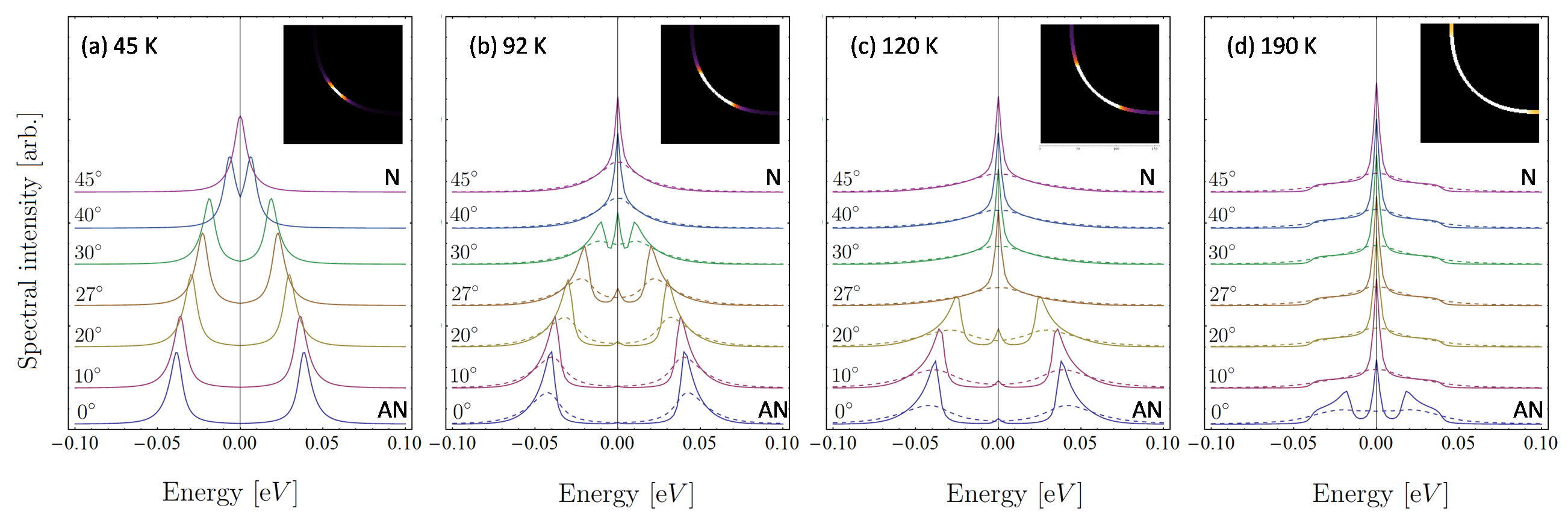}
\caption{(Color online) Spectral function plots at the Fermi energy
$A(\vec{k_F},E)$ as a function of angle at different temperatures
$T=$45, 92, 120, 190\,K. Plain lines: Spectral function calculated
for $\Gamma=$2\,meV; Dashed lines: Spectral function calculated for
$\Gamma=k_B\,T$. Note that, contrary to a BCS picture, it is the
excited pairs that dissociate giving rise to the Fermi arcs (peaks
at the Fermi level). Insets : calculated Fermi arcs in the reduced
Brillouin zone.} \label{Fig_arcs}
\end{figure*}

We start with the low-temperature SC state where all pairs belong to
the condensate. An essential concept of the model is that, upon
rising temperature, pairs are excited out of the condensate without
pair-breaking -- a highly non-BCS feature.

Taking the relevant Bose-Einstein statistics with $\mu = 0$, and
assuming $d$-wave pairing, the condensate angular density is given
by\,:
\begin{eqnarray}
&n_{oc}(T,\theta)& = n_0 -\mathcal{A}(T,\theta)\,\int_{\Delta_p
{\rm cos}(2\theta)+\delta}^{\infty}\,{\rm d}\Delta_i\,P_0(\Delta_i) \hskip 0.1 cm \null \nonumber\\
&\null& \times \,  f_B(\Delta_i-\Delta_p\,{\rm cos}(2\theta), T)\ ,
\label{Noc}
\end{eqnarray}
where $f_B(E,T)=(e^\frac{E}{k_BT}-1)^{-1}$ is the Bose-Einstein
distribution, $\delta$ is a low-energy cut-off
\cite{SciTech_Sacks2015} and $\mathcal{A}(T,\theta)$ is a
normalization factor to be discussed below. The constant $n_0$,
directly proportional to the doping value, is assumed to be
independent of $\theta$.

The integrated condensate density, $\int
d\theta\,n_{oc}(T,\theta)/(2\,\pi)$, shown in Fig.\,\ref{Fig_noc},
is a monotonically decreasing function of temperature (lower curve),
due to pair excitations, and vanishes at $T_c$ as expected. Note
that in this temperature range, the antinodal gap $\Delta_p(T)$
(upper curve), reflecting the total number of pairs, is practically
constant up to $T_c$, in agreement with experiment. However for
higher temperatures, $T > T_c$ the gap $\Delta_p(T)$ markedly
decreases to finally vanish at $T^*$ as a result of pair
dissociation\,\cite{EPJB_Sacks2016}.

The normalization factor $\mathcal{A}(T,\theta)$ is determined using
appropriate boundary conditions.  We assume that, even at finite
temperature (below $T_c$), the condensate density remains uniform\,:
$n_{oc}(T,\theta) = n_{oc}(T)$. Furthermore, the temperature
dependence of the antinodal gap $\Delta_p(T)$ (Fig.\,\ref{Fig_noc})
is taken throughout this work as the BCS function, however with the
ratio $\Delta_p(0)/k_B T^* = 2.2$ compatible with the gap equation
(\ref{Delta_ki}). Finally, imposing $n_{oc}(T_c)$ =0, we obtain a
self-consistent form for $\mathcal{A}(T,\theta)$ with the
quadrature\,:
\begin{eqnarray}
&\mathcal{A}(T,\theta)^{-1}& = (1-n_{oc}(T))^{-1} \hskip 0.2 cm  \nonumber\\
&\null& \times \, \int_{\Delta_p {\rm
cos}(2\theta)+\delta}^{\infty}\,{\rm d}\Delta_i\,P_0(\Delta_i) \nonumber \\
&\null& \hskip 10 mm \times \, f_B(\Delta_i-\Delta_p\,{\rm
cos}(2\theta), T) \ .\label{AofToftheta}
\end{eqnarray}
Note that the latter self-consistent equation implies the constraint
of particle conservation at all temperatures.

Let us now consider the excited-pair and dissociated-pair densities.
The latter dissociation phenomenon occurs if the pair binding energy
is typically small compared to the thermal energy. As in BCS theory,
this process is governed by the Fermi-Dirac distribution
$f(E,T)=(e^\frac{(E-\mu)}{k_BT}+1)^{-1}$ giving rise to the $\left[
1-\tanh\left(\frac{E}{k_BT}\right)\right]$ factor in the number of
dissociated pairs:
\begin{eqnarray}
&n_{diss}(T,\theta)& =\mathcal{A}(T,\theta) \,\int_{\Delta_p
{\rm cos}(2\theta)+\delta}^{\infty}\,{\rm d}\Delta_i\,P_0(\Delta_i) \hskip 0.1 cm \null \nonumber\\
&\null& \times \, f_B(\Delta_i-\Delta_p\,{\rm
cos}(2\theta),T) \hskip 0.1 cm \null \nonumber\\
&\null& \times
\,\left[1-\tanh\left(\frac{\Delta^i}{kT}\right)\right]\ .
\label{Ndiss}
\end{eqnarray}

In a similar way, one can write the excited pair density\,:
\begin{eqnarray}
&n_{ex}(T,\theta)& =\mathcal{A}(T,\theta) \,\int_{\Delta_p
{\rm cos}(2\theta)+\delta}^{\infty}\,{\rm d}\Delta_i\, P_0(\Delta_i) \hskip 0.1 cm \null \nonumber\\
&\null& \times \, f_B(\Delta_i-\Delta_p\,{\rm cos}(2\theta),T)\hskip 0.1 cm \null \nonumber\\
&\null& \times \, \tanh\left(\frac{\Delta^i}{kT}\right)\
.\label{Nex}
\end{eqnarray}

The three densities must follow the sum rule\,:
\begin{equation}
n_{oc}(T)+n_{ex}(T,\theta)+n_{diss}(T,\theta)=n_0\ ,
\end{equation}
which can be verified by inspection. Furthermore, they inherently
capture the physical properties of the cuprate phase diagram, from
$T=0$ through $T_c$ up to $T^*$. It is remarquable that both
Bose-Einstein and Fermi-Dirac statistics appear in $n_{ex}$ and
$n_{diss}$ -- a direct consequence of the composite nature of the
pairons.

\vskip 2mm
{\it Spectral function}.
\vskip 2mm

Similarly as for the DOS \cite{EPJB_Sacks2016}, the spectral
function $A(\vec{k},E)$ can be expressed as a sum of three terms,
the condensate spectral function  $A_{cond}(\vec{k},E)$, the excited
pairs contribution $A_{ex}(\vec{k},E)$ and finally the dissociated
pairs term $A_{diss}(\vec{k},E)$. The first term is essentially
determined by the number of condensed pairs with energy $\Delta_p$,
associated with the quasiparticles
$E_{\vec{k}}=\sqrt{\epsilon_{\vec{k}}^2+\Delta_{\vec{k}}^2}$, where
$\Delta_k$ is the condensate gap function\cite{SciTech_Sacks2015}\,:
\begin{equation}
A_{cond}(\vec{k},E)=\frac{-1}{\pi}\Im
m{\frac{n_{oc}(T)}{E-E_k+i\Gamma}}\ .\label{Spec_fct}
\end{equation}
$\Gamma$ is the standard parameter describing energy broadening.

In the latter equation we have neglected the coherence factors for
occupied versus unoccupied states, which corresponds to the spectral
function for excited states (both hole and electron-type). In this
case, as shown by Schrieffer \cite{RMP_Schrieffer1964}, for
symmetric bands at the Fermi level, the coherence factors disappear.
Since we are concerned here with energies very close to the Fermi
level, where the particle/hole asymmetry is small (i.e. with no
additional self-energy), equation \ref{Spec_fct} provides a good
approximation for the EDCs.

The excited-pair term of the spectral function results from thermal
excitations of pairons out of the condensate:

\begin{eqnarray}
&A_{ex}(\vec{k},E)& =\frac{-\mathcal{A}(T,\theta)}{\pi}\, \Im
m{\int\frac{d\Delta_i\,P_0(\Delta_i)}{E-E_{\vec{k}}^i+i\Gamma}} \hskip 0.1 cm \null \nonumber\\
&\null& \times
f_B(\Delta^i-\Delta_p(T),T)\tanh\left(\frac{\Delta^i}{kT}\right)\ .
\label{A_ex}
\end{eqnarray}
Finally, the dissociated pair term is caused by the thermal
dissociation of Cooper pairs into normal fermions of energy
$\epsilon_{\vec{k}}$. It has the simple expression:
\begin{equation}
A_{diss}(\vec{k},E)=\frac{-\mathcal{A}(T,\theta)}{\pi}\Im
m{\frac{n_{diss}(T,\theta)}{E-\epsilon_{\vec{k}}+i\Gamma}}\ .
\end{equation}

The total spectral function at the Fermi energy $A(\vec{k}_F,E)$,
comparable to the measured EDC, is shown in Fig.\ref{Fig_arcs} as a
function of angle for four different temperatures, ranging from low
$T$ up to a temperature close to $T^*$. The corresponding Fermi
surface is indicated by the inset in each panel. Two values of the
broadening parameter were considered. First a small value,
$\Gamma=$2\,meV to highlight the intrinsic spectral weight (solid
lines). Secondly, choosing $\Gamma=k_BT$ accounts for the thermal
broadening (Fig.\ref{Fig_arcs}, dashed lines) present in
experiments.

At low temperature, $T=45$\,K (which is well below
$T^*\approx$\,200\,K), two well-defined Bogoliubov peaks are clearly
visible in the spectra (Fig.\ref{Fig_arcs}, panel (a)). The latter
is maximum in the AN direction, at the energy $\pm \Delta_p$, and
decreases as a function of angle to vanish in the N direction in
agreement with $d$-wave symmetry\,: $\Delta_p\,{\rm cos}(2 \theta)$.
However, in the vicinity of the node, the spectra exhibit a peak at
$E=0$ which originates from dissociated pairs. Thus, at low
temperature the spectral function $A(\vec{k}_F,E=0)$ is zero along
the Fermi surface except around the N direction where a tiny arc is
revealed (insert of Fig.\,\ref{Fig_arcs}\,(a)). This arc of normal
states reduces to a `Fermi point' at zero temperature, in agreement
with experiments \cite{NatPhys_Hashimoto2014}.

\vskip 1mm

At the critical temperature (Fig.\ref{Fig_arcs}, panel (b)), the gap
closes well before the node at the critical angle $\theta_c$ above
which the peaks remain at the Fermi level. This effect constitutes
the critical Fermi arc which, as will be explained below, results
from pair breaking concomitant with thermally induced quasiparticle
excitations. Even above $T_c$, panel (c), {\it Bogoliubov coherence
peaks are still present in the AN direction}, due to excited pairs
in the pseudogap state. As the temperature continues to increase,
the Fermi arc progressively expands and finally the full Fermi
surface is almost recovered at $T=$190\,K, which is close to
$T^*\approx$\,200\,K (panel (d)).

\vskip 4mm {\it Comparison to ARPES experiments} \vskip 2mm

\begin{figure}[t]
\centering \vbox to 7.4 cm{
\includegraphics[width=8.8 cm]{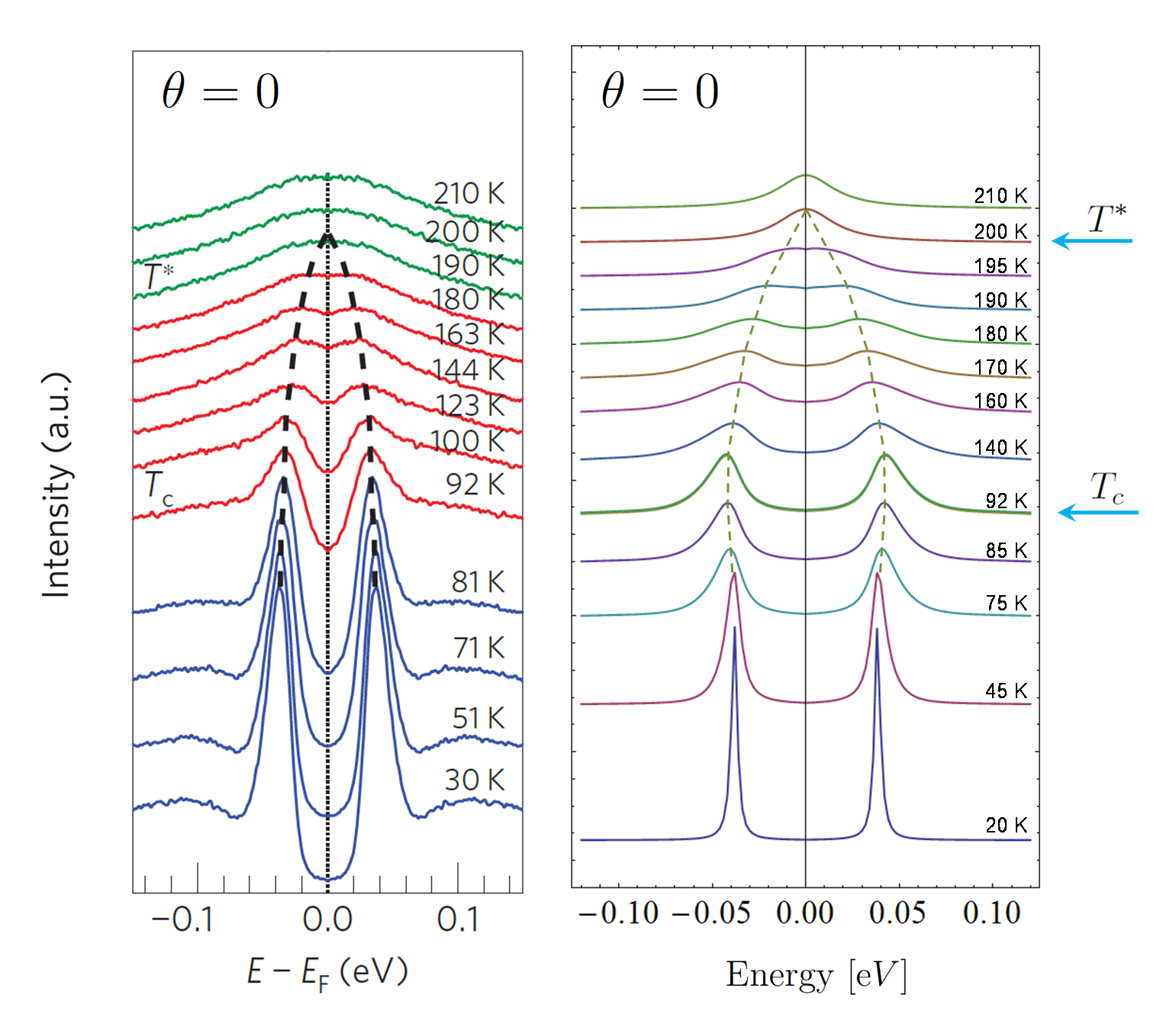}}
\caption{Left panel\,: Measured ARPES EDC spectra (symmetrized) at
the Fermi surface as a function of temperature in the anti-nodal
direction (data from Hashimoto et al.\cite{NatPhys_Hashimoto2014}).
Right panel\,: Calculated Spectral function at the Fermi surface
$A({\vec k}_F,E)$ as a function of temperature in the antinodal
direction. The broadening parameter is $\Gamma = k_B\,T$. Note that
the incoherent background of the experimental spectrum is not
accounted for.}\label{Fig_Hashi_antinode}
\end{figure}

\begin{figure}
\centering \vbox to 5.8 cm{
\includegraphics[width=8.6 cm]{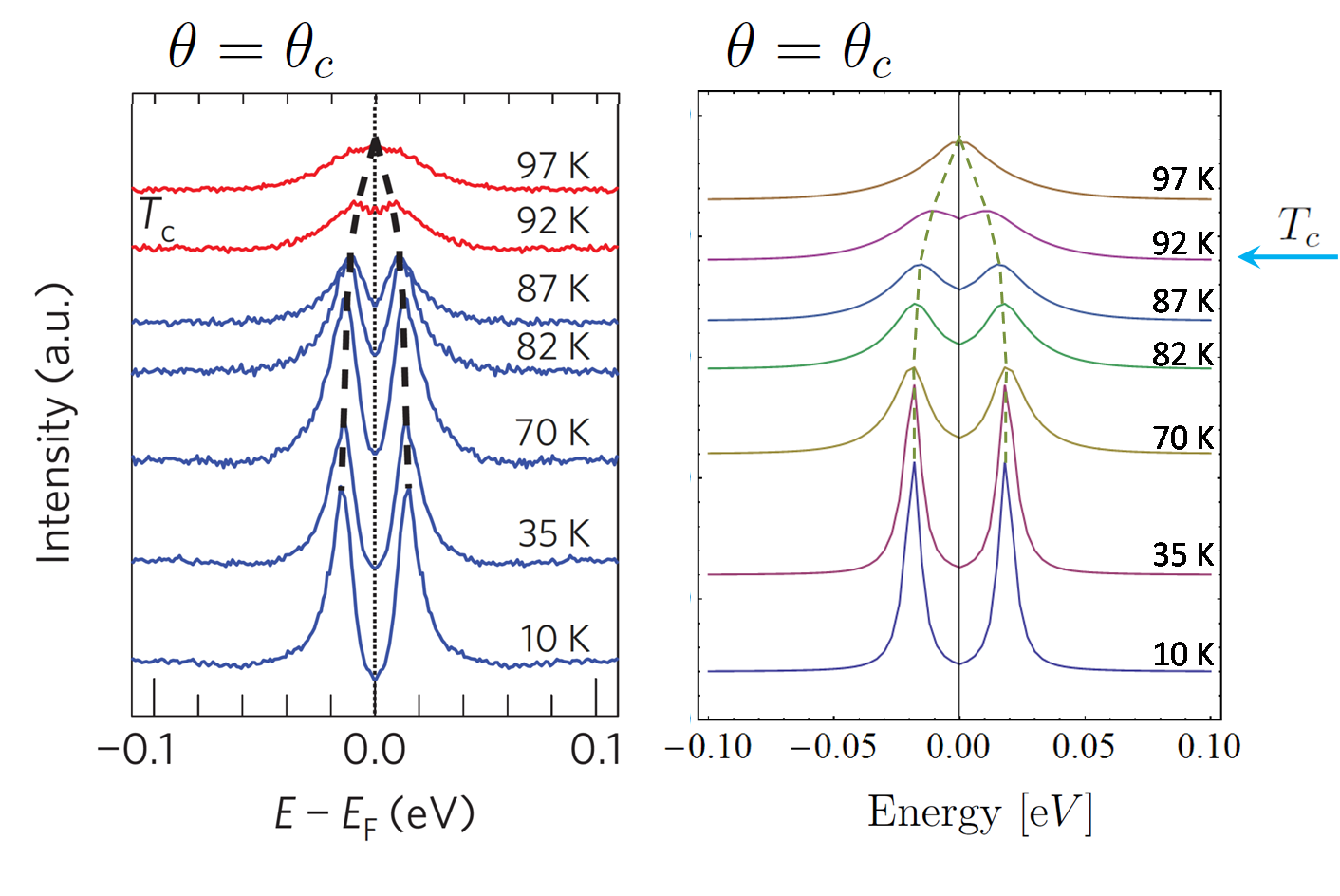}}
\caption{Left panel\,: Measured ARPES EDC spectra (symmetrized) at
the Fermi surface as a function of temperature near the node
($\theta=$31$^\circ$) (data from Hashimoto et
al.\cite{NatPhys_Hashimoto2014}). Right panel\,: Calculated Spectral
function at the Fermi surface $A(\vec{k_F},E)$ as a function of
temperature near the node. The broadening parameter is $\Gamma =
k_B\,T$.}\label{Fig_Hashi_node}
\end{figure}

Hashimoto et al.\cite{NatPhys_Hashimoto2014} have done extensive
ARPES studies of slightly underdoped
Bi$_2$Sr$_2$CaCu$_2$O$_{8+\delta}$ ($T_c = 92$\,K). In
Fig.\ref{Fig_Hashi_antinode} and Fig.\ref{Fig_Hashi_node}, we
compare the temperature dependence of the calculated EDC spectra at
the anti-node ($\theta = 0$) and near the node at the critical angle
($\theta_c = 31^\circ$) to the experimental spectra reported in
\cite{NatPhys_Hashimoto2014}. The overall similarity between the two
sets of spectra is striking. As discussed previously, in
Fig.\ref{Fig_Hashi_antinode} one observes clear Bogoliubov peaks at
$\sim 40$\,meV that remain virtually constant in energy up to $T_c$
and then slowly close and vanish at the higher temperature $T^*$. On
the contrary, at the critical angle nearer the node
(Fig.\ref{Fig_Hashi_node}) the closing of the gap, initially of
smaller value $\sim 19$\,meV, is at $T_c$. This Fermi arc formation
is directly seen in Fig.\ref{Fig_arcs}, panel (b). Further ARPES
data confirms that the closing temperature of the Bogoliubov peaks
is indeed a monotonic function of angle at the Fermi surface.

The temperature-dependent Fermi arc is thus governed by a single
mechanism. Consider the natural hypothesis that each Cooper pair, as
in the BCS theory \cite{PR_BCS1957}, is subject to quasiparticle
excitations as a function of temperature. Consequently, the gap
amplitude $\Delta(\theta)$ for a given pair must follow the BCS-type
temperature dependence and close at a temperature such that
$\Delta(\theta)/k_B\,T\sim C$, where $C$ is a constant of order 2. A
simple extension of the pairon gap equation (\ref{Delta_ki}) to
finite temperature gives this result. From the available ARPES and
tunneling data we find a good agreement using $C=$2.2. We note that
this value is slightly larger than the BCS value, $C_{BCS} \simeq
1.76$, indicating strong coupling.

\begin{figure}[t!]
\centering \vbox to 5.6 cm{
\includegraphics[width=6.8 cm]{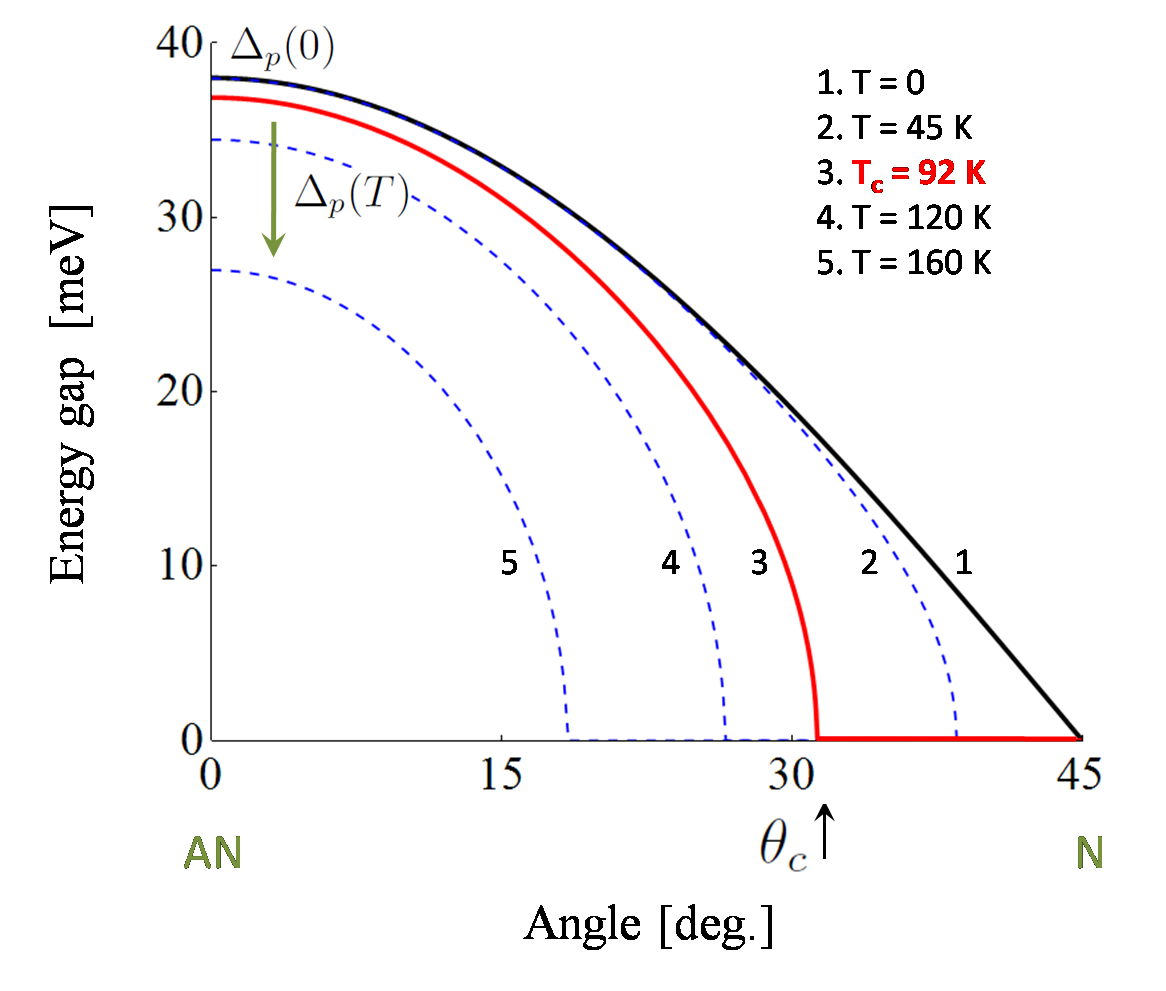}}
\caption{(Color online) Theoretical angular and temperature
dependence of the gap function (Bogoliubov peak),
$\Delta_p(T,\theta)$, illustrating the Fermi-arc formation near the
node (N). For very low temperatures, only a tiny Fermi arc exists
around the node and the gap is essentially $d$-wave. At $T_c$ the
Fermi arc extends to the critical angle $\theta_c$ while the
antinodal gap (AN) has hardly decreased (from underdoped to
optimally doped cases). Above $T_c$ the Fermi arcs continue to grow
but now the antinodal gap decreases concomitantly.
}\label{Fig_delta_theta}
\end{figure}

The overall interpretation of the angular and temperature evolution
of the ARPES EDCs is illustrated in Fig.\,\ref{Fig_delta_theta}. At
zero temperature, the condensate gap $\Delta_{\vec{k}}$ has the
standard $d$-wave symmetry, with the angular dependence
$\cos(2\theta)$, closing in the N direction, the Fermi point
observed in ARPES measurements.

As the temperature rises, some pairons are excited out of the
condensate. In the AN direction, their concentration $n_{ex}(T,
\theta = 0)$ is determined by Bose statistics, however their binding
energy decreases with $\theta$ due to the $d$-wave symmetry and, for
a given angle, $\Delta_p(T,\theta)$ decreases according to the BCS
function. Therefore all the excited pairons with energy less than
$2.2\,k_B\,T$ will dissociate, leading to the Fermi arc of normal
states in the region near the node (between  45$^{\circ}$ and the
intercept between the $\Delta_p(T,\theta)$ curve with the abscissa
in Fig.\,\ref{Fig_delta_theta}).

We stress that this nodal pair dissociation is due to quasiparticule
excitations driven by the Fermi-Dirac statistics, as given
explicitly in Eqs.(\ref{Ndiss},\,\ref{Nex}). At the critical
temperature, the Fermi arc has an angular width $\theta_c$ to
$\frac{\pi}{2}-\theta_c$ where $\theta_c$ satisfies the relation\,:
\begin{equation}
\Delta_p(0)\times\cos(2\theta_c)=2.2\,k_B\,T_c\ .
\label{Eq_Deltap_Tc}
\end{equation}

While weakly bound pairs first dissociate near the node, pairs
persist in the antinode and are progressively excited out of the
condensate up to the critical temperature where $\Delta_p(T_c,\theta
=0)$ has hardly decreased and where $n_{oc}(T_c)=0$ (see
Fig.\,\ref{Fig_delta_theta}, red curve). Note that the pairons in
the condensate do not directly contribute to the Fermi arcs, the
consequence of our assumption that their concentration $n_{oc}(T)$
is independent of angle.

Above $T_c$, as the arc continues to extend away from the node, the
gap progressively closes in the antinode as more and more pairs are
being dissociated into quasiparticles. Finally, all pairs are
dissociated near $T^*$, where the full Fermi surface is recovered.
As discussed previously, this temperature is given empirically by
the relation\,:
\begin{equation}
\Delta_p(0)=2.2\,k_BT^* \ ,
\end{equation}
regardless of the doping. Therefore, using (\ref{Eq_Deltap_Tc}), we
have the important relation
\begin{equation}
\cos(2\theta_c) \simeq \frac{T_c}{T^*} \ , \label{ratio}
\end{equation}

This equation should be valid for all doping values in the phase
diagram. Thus the critical angle should satisfy\,: $\cos(2\theta_c)
= f(p)$, where $f(p)$ is a unique function of the carrier
concentration.

In the underdoped limit ($T^* \gg T_c$) the critical angle is close
to $\pi/4$ indicating very small nodal Fermi arcs at $T_c$. Then the
antinodal boson character dominates the phase transition. To the
contrary, on the overdoped side ($T^* \sim T_c$) the Fermi arcs grow
very rapidly with temperature indicating the coexistence of fermion
and boson excitations, in good agreement with experiments
\cite{PhilMag_Kaminski2015,PRB_Ideda2012}. Thus, the cuprate
phenomenology is in fact continuous as a function of doping while
the simple relation (\ref{ratio}) implies that the superconducting
order, the pseudogap state and the Fermi arcs are intimately linked.

\vskip 2mm {\it Conclusion} \vskip 2mm

We have shown that ARPES measurements in cuprates can be understood
in the context of a Bose-Einstein condensation of preformed pairs.
The nature of these composite bosons, pairs of holes in their
antiferromagnetic environment, or pairons, naturally imposes a
$d$-wave symmetry of the order parameter and gives the correct
energy scale of the phase transition, the exchange energy $J$.

Clear Bogoliubov peaks in ARPES and corresponding pseudogap in
tunneling reveal the preformed pairs which, due to their strong
binding energy, persist in the system at all temperatures below
$T^*$. Below $T_c$, boson excitations dominate in the antinodal
direction (strong pairing) while fermion excitations dominate near
the node (weak pairing), giving rise to the Fermi arcs. Above $T_c$,
the preformed pairs dissociate continuously, while the Fermi arcs
grow at a rate which is doping dependent. Finally, the complete
Fermi surface is recovered at $T^*$.

In conclusion, the superconducting state, the pseudogap state and
the Fermi arcs are tied together as a consequence of a unique
phenomenon, the pairons and their excitations.

\vskip 2mm {\it Acknowledgements} \vskip 2mm

We gratefully thank N.\,Miyakawa (Tokyo University of Science),
A.\,Fujimori (University of Tokyo) and H.\,Eisaki (AIST Tsukuba) for
stimulating discussions.

\end{document}